# A New Framework for Understanding Supersolidity in $^4$He

David S. Weiss

Physics Department, The Pennsylvania State University, University Park, PA 16802, USA

Recent experiments have detected $^4$He supersolidity, but the observed transition is rather unusual. In this paper, we describe a supersolid as a normal lattice suffused by a Bose gas. With this approach, we can explain the central aspects of the recent observations as well as previous experimental results. We conclude that detailed balance is violated in part of the $^4$He phase diagram, so that even in steady state contact with a thermal reservoir, the material is not in equilibrium.

The possibility that quantum solids exhibit supersolidity has been debated for 45 years[1,2]. The experimental search for the phenomenon culminated in the recent observations of Kim and Chan that the moment of inertia of solid $^4$He decreases at low temperature, both in porous media[3] and in bulk[4]. These experiments strongly suggest a quantum phase transition to a supersolid state, especially because the effect is absent in solid $^3$He or when the annulus of their sample is blocked. However, the behavior of the transition is unexpected. It is gradual as a function of temperature, instead of sharp. The non-classical inertial effect starts to disappear at a very low critical velocity. When the supersolid is slightly doped with $^3$He, the reduction in inertia decreases, but the apparent transition temperature actually increases.

In this paper, we present a new framework for understanding supersolidity in $^4$He. While consistent with earlier theoretical conceptions[5,6], our framework allows for an interpretation of all available experimental data and for prediction of new phenomena. Our goal here is to explain the basic physics of the supersolid, setting aside many technical details. Briefly, we describe the supersolid as a normal lattice suffused by a Bose gas of $^4$He quasiparticles (not vacancies). Below $T_c$, the critical temperature where the Bose gas starts to form a Bose-Einstein condensate (BEC), intra-gas and phonon scattering interact in a way that disrupts detailed balance and leads to non-Bose distributions of the gas. For a range of temperatures the supersolid exists in a steady state that is not thermal equilibrium. As the phonon energy density decreases with temperature, a sizable BEC forms and supersolid effects can be observed. The supersolid only attains thermal equilibrium below a new temperature, $T_s$, the physics of which we will describe.

Supersolidity requires a quantum solid[7], where large tunneling rates lead to quantum fluctuations in the number of atoms at each lattice site. The existence of small number fluctuations allows phase coherence to extend over multiple lattice sites. The part of the wavefunction that is phase coherent across lattice sites can be parameterized as the supersolid component[5,6]. Equivalently, the phase coherence can be associated with delocalized quasiparticles, and this is the description we use. To understand inertial effects in a supersolid from this perspective, first consider an individual quasiparticle delocalized across many lattice sites. Without collisions, quasiparticle wavefunctions do not move when the lattice is accelerated[8,9]. However, collisions can disrupt the quasiparticle wavefunctions and cause them to be dragged along with the lattice. The macroscopic state occupation of a BEC makes its wavefunction stable against collisions. It retains its form even in the extreme case that a large fraction of the constituent particles are scattered out of the state. Thus a BEC, like a lone quasiparticle without collisions, is not dragged by the lattice. Changes in the moment of inertia of a supersolid are proportional to the number of quasiparticles in the BEC (below the critical velocity).

It has been widely supposed that the quasiparticle responsible for supersolidity would be vacancies, the delocalized absence of $^4$He atoms[1,2,10,11]. Because Kim and Chan observe that the moment of inertia at low temperature goes down, not up, we conclude that the quasiparticles are not vacancies, but delocalized $^4$He atoms. If the dominant quasiparticles were vacancies, the supersolid would be like a quantum Swiss cheese, where the holes don't move when the cheese does. That would require some of the atoms to move faster than they otherwise would, which takes more energy than if the holes moved, resulting in a higher moment of inertia[12].

In seeking to justify why the supersolid, like a superfluid, can be described as a two-component object, it may, be instructive to consider the Bose-Hubbard model[13], which differs from the real quantum solid in that the periodic potential is externally imposed. When tunneling energies dominate on-site repulsion, the system is a superfluid (SF), and in the opposite limit it is a Mott insulator (MI). Near the SF-MI transition, consider the system on interatomic scattering timescales, which are long compared to intersite tunnelling rates. Then the system is partly like an MI, in the sense that there is a site occupancy, which can be less than or equal to one atom per site, below which there are no fluctuations. It is also partly like an SF, with small number fluctuations and long range phase coherence. We assume in this paper that it is possible to divide such a system into an MI-like part with a fractional atom at each site, and an SF-like part consisting of $N_q$ quasiparticles. Presumably, $N_q$ and $m_{eff}$ depend on the proximity of the quantum phase transition and the ratio of $N$ to the number of lattice sites. Since tunneling energies greatly exceed relevant thermal energies, $N_q$ and $m_{eff}$ would be expected to be temperature independent. In solid $^4$He, $N_q$ and $m_{eff}$ depend only on pressure and details of the interatomic potentials. Rigourous justification of this picture requires further theoretical work.

Three scattering processes affect the Bose gas component of the quantum solid: intra-gas scattering, wall collisions, and phonon scattering. The first two are straightforward, and sufficient to make the gas thermally equilibrate. Phonon scattering deserves special consideration. For definiteness, consider $^4$He solid in an annular container with a radius $R$=1 cm, as in Kim and Chan's experiment[4]. The possible phonon energies are given by $E_{ph}(n)= n\hbar kv$, where $v$ is the sound velocity (~600 m/s for solid $^4$He), k=1/$R$ is the wavenumber of the lowest energy excitation, and $n$ is a positive integer. The allowed energy states of the Bose gas, ignoring mean field interactions, are given by $E_q(n)=m^2(\hbar k)^2/2m_{eff}$, where k is the same as for the phonons and $m$ is a non-negative integer. The ratio $Q=E_{ph}(1)/ E_q(1)$ is huge, $2\times10^7$. Ignoring energy level broadening for the moment, a quasiparticle in a low energy state cannot absorb a phonon and conserve both energy and quasi-momentum. Phonon scattering is allowed only when adjacent quasiparticle levels are separated by the minimum phonon energy, i.e. above $m\approx10^7$, which corresponds to 2.4 K, hotter than where supersolid effects are observed. In fact, energy level broadening is important, because intra-gas scattering rates are much higher than phonon frequencies. So phonons presumably do play a role in thermalizing the quasiparticles.

A completely new phonon-quasiparticle scattering process can occur below $T_c$, when the ground state is macroscopically occupied. A macroscopically occupied state with $M$ quasiparticles can be described by the wavefunction $\sqrt{M}e^{im\mathbf{k}\cdot\mathbf{r}}$ [14]. For the BEC, $M=N_c$ and $m$=0. Because they are coherent, ground state quasiparticles can collectively absorb a single phonon, and in so doing make a transition to a macroscopically occupied excited state. Energy is conserved when an $E_{ph}(n)$ phonon is collectively absorbed by $Q/n$ ground state quasiparticles, leaving them with the wavefunction $\sqrt{\frac{Q}{n}}e^{in\mathbf{k}\cdot\mathbf{r}}$. This transition satisfies the Bragg condition $\mathbf{k_{ph}}(n)+\mathbf{k_q}(0)= \mathbf{k_q}(n)$ (i.e., $\mathbf{k_{ph}}(n)= \mathbf{k_q}(n)$), which is the same as for the individual quasiparticles. The transition violates quasi-momentum conservation by a factor of $Q/n$, but it is allowed by an unanticipated loophole in the law. For lone particle scattering, quasi-momentum conservation results from the need to satisfy the Bragg condition[15]. It is the latter that is fundamental. Of course, when $Q/n$ particles collectively absorb a phonon, real momentum must be conserved by recoil of the lattice, as in the Mössbauer effect.

After groups of BEC quasiparticles have collectively phonon scattered to a higher momentum state, the other scattering processes break up the coherent groups, spreading them among momentum states. This removes the return path to the ground state via collective phonon scattering. The other scattering processes will then tend to return the quasiparticles to the ground state. The irreversibility of collective phonon scattering in the presence of other scattering leads to a one way loop in population transfer, as outlined in Fig. 1. Detailed balance is lost, and with it the expectation that the quasiparticles obey the Bose distribution. There will be fewer atoms in the BEC than the Bose distribution would predict. Although steady state, non-equilibrium processes are well known in physics[16], the supersolid may be the first example of a material in steady state contact with a heat bath that does not internally thermalize.

We can model the above loop process with a simple rate equation,

$\frac{dN_c}{dt} = -\Gamma_p(T)N_c + \Gamma_c(N_c, T)$, with the constraint that $N_c$ cannot be greater than what the

Bose distribution would predict, $N_{c0} = N_q\left[1 - \left(\frac{T}{T_c}\right)^{\frac{3}{2}}\right]$. $\Gamma_p$ is the total irreversible loss rate from the ground state, and $\Gamma_c$ is the average rate at which atoms are returned to the ground state by the other scattering processes. The phonon energy density of a perfect crystal well below the Debye temperature is proportional to $T^4$ [17]. Because excitation from the ground state is the loss rate limiting step, $\Gamma_p = \alpha\, T^4$, where $\alpha$ is the sum of collective phonon scattering matrix elements.

The temperature dependence of $\Gamma_c$ will be dominated by the temperature dependence of quasiparticle scattering rates (which are proportional to the average quasiparticle velocity), so $\Gamma_c \approx \beta\sqrt{T}$. The derivation of the functional form of $\beta$ is a complicated non-equilibrium kinetics problem. It must, however, lie between two extremes. The fastest possible approach to equilibrium would be if $\beta$ were proportional to the constant $N_q$, so that the rate does not decrease as equilibrium is approached. The slowest would be if it were proportional to the number of non-equilibrated quasiparticles, $N_{c0}-N_c$, so that the rate approaches zero near equilibrium. For $\beta=\beta_1 N_q$, $N_c = \dfrac{N_q}{\left(\dfrac{T}{T_{sa}}\right)^{3.5}}$ where $T_{sa}=(\beta_1/\alpha)^{1/3.5}$, with the constraint that $N_c \leq N_{c0}(T)$. For $\beta=\beta_2(N_{c0}-N_c)$, $N_c = \dfrac{N_{c0}(T)}{1+\left(\dfrac{T}{T_{sb}}\right)^{3.5}}$, where $T_{sb}=(\beta_2/\alpha)^{1/3.5}$. In Fig. 2, experimental data for $N_c$ at different pressures[4] are compared to these two limiting functions. In each case, the only free parameters are $N_q$, $T_{sa}$ or $T_{sb}$, and $T_c$, where the dependence on $T_c$ is very weak. The particular values of $T_s$ and $N_c$ are sample dependent (even at fixed pressure, presumably because of inconsistently grown crystals[4]), as is the location of the data between the two limiting theory curves, but all the data are in the range consistent with theory. The 26 and 41 bar curves are completely insensitive to $T_c$, as long as it is larger than ~1 K, while the 65 bar curve fits somewhat better with $T_c \approx 0.4$ K. It is apparent that $T_c$ is greater than the highest temperature at which non-classical rotational inertia is observed. We conclude that detailed balance is violated below $T_c$ and above $T_s$, which we define as the point that the data starts to coincide with $N_{c0}$. In that range, the quasiparticle gas does not conform to the Bose distribution. With more reproducible crystal growth, it may be possible to extract from $N_c(T)$ detailed information about scattering processes in the unequilibrated supersolid.

The supersolid's very low critical velocity, $v_c$, is another result of collective phonon scattering, by a process we will describe elsewhere. For present purposes, we speculate that $v_c$ corresponds to the BEC having one unit of circulation. Then $m_{eff}=\hbar/Rv_c$. In the Kim and Chan experiments, $m_{eff}$ varies from $0.1 m_{He}$ near the melting curve to ~$2 m_{He}$ at P=65 bar. Measured values for $v_c$ and the non-classical inertial mass density, $\rho_{NC}$, allow us to calculate $T_c$, $T_c = 2\pi\hbar^{\frac{1}{3}} \left( \frac{\rho_{NC}}{2.612} \right)^{\frac{2}{3}} (2\pi R v_c)^{\frac{5}{3}}$. Since $\rho_{NC}$ and $v_c$ depend somewhat on individual crystals, our results for $T_c$ are not definitive[3,4]. But we find that $T_c$ is always higher than the temperatures at which non-classical inertial effects are seen and tends to decrease with increasing pressure, both of which are consistent with the analysis of the phase transition above. We can use these values for $T_c$, along with measurements of $T_s$, to suggest a new phase diagram for $^4$He, shown in Fig. 3. Because of the unequilibrated supersolid phase, the normal solid to supersolid transition cannot be characterized in conventional terms. The phase diagram suggests that at higher pressures, $T_c$ might drop below $T_s$, which would make for a typical second order phase transition.

Our understanding of the $^4$He supersolid explains the observation that slight doping with $^3$He reduces the supersolid fraction, yet appears to increase the transition temperature[3]. With a smaller supersolid fraction, assuming a comparable $m_{eff}$, $T_c$ is lower. But $T_c$ is not actually observed. Rather, $T_s$ is observed, along with the directly related maximum temperature at which non-classical inertial effects can be seen. The increase in $T_s$ due to $^3$He doping would occur, for instance, if the presence of $^3$He atoms decreases the matrix element for quasiparticle-phonon scattering. Similarly, we can understand why the supersolid's phase transition is so different from the superfluid's. While the same disrupted equilibrium process should occur in the fluid, it would not be seen if $T_s > T_c$, as would hold if the phonon scattering rates were much smaller. Significantly smaller phonon scattering rates are also consistent with the much higher $v_c$ in the superfluid than the supersolid.

Most attempts to observe $^4$He supersolidity have been searches for non-classical mass flow[18]. Using our framework, we can understand why no effect is observed. The relative fraction of the normal lattice and Bose gas is fixed at a given pressure. The Bose gas can move through the normal lattice only if its local average density does not change, as in rotational flow. Since the normal lattice cannot flow between chambers, neither can the Bose gas. Goodkind has observed evidence for a novel transport mechanism in the vicinity of the unequilibrated supersolid[19]. This mechanism, which is absent at lower temperatures, would be explained within our framework if it was associated with the normal quasiparticle Bose gas. In our picture, there are three transport carriers: lattice phonons; the normal quasiparticle Bose gas at temperatures from $T_s$ to above $T_c$ (as long as $^4$He is still a quantum solid); and the BEC below $T_s$ and to a degree in the unequilibrated supersolid regime.

Along with modifications in transport and inertial effects, phase transitions are often manifested by changes in heat capacity, $C_v$. In the unequilibrated supersolid, effects near $T_c$ are absent, since the BEC's contribution to $C_v$ should mimic $N_c$ as a function of temperature. Thus the most easily observable effect is near $T_s$, where $dN_c/dT$ is largest. If the unequilibrated supersolid region does disappear at higher pressure, a heat capacity change could be seen at $T_c$.

In conclusion, we have presented a physical picture of supersolid $^4$He that accords with all experiments to date. Future experiments that could test this picture include direct observation of collective phonon scattering, modification of the transition shape at higher pressures and/or different $^3$He doping, and heat capacity and transport studies. In its macroscopic occupation of the ground state, a supersolid is similar to a superfluid[20] and a dilute BEC gas[21]. The normal solid to supersolid transition, however, seems unique. Two different lines of arguments have led us to conclude that $T_c$ is greater than the temperatures at which supersolid effects are observed. Collective phonon scattering, which becomes irreversible in the presence of other scattering, explains the gradual manifestation of supersolidity. It also implies that in part of the $^4$He phase diagram, the supersolid is not internally thermalized.

The author thanks M. Chan, J. Jain, K. Gibble, J. Banavar, M. Cole, A. Mizel, P. Schiffer, V. Crespi, and P. Lammert for useful discussions, E. Kim and M. Chan for sharing their data, and Penn State, the NSF, DARPA, and the Packard Foundation for their support.

[8] Very similar physics has been demonstrated with delocalized atoms in an accelerating optical lattice[9]. When the atoms are accelerated to the edge of the Brillouin zone, they Bragg scatter off the lattice and catch up with it (a Bloch oscillation). In the Kim and Chan experiments, velocities are orders of magnitude below the Brillouin zone edge, so there is no Bloch oscillation.

[9] M. B. Dahan, E. Peik, J. Reichel, Y. Castin, C. Salomon, *Phys. Rev. Lett.* **76**, 4506 (1996).

[10] M. J. Bijlsma, H. T. C. Stoof, *Phys. Rev. B* **56**, 14361 (1997).

[11] C. A. Burns, *Physica B* **253**, 180 (1998).

[12] Alternately, find the moment of inertia of a disk with a hole in it by assigning a negative moment of inertia to the hole, and then keep the hole fixed.

[13] M. P. A. Fisher, P. B. Weichman, G. Grinstein, D. S. Fisher, *Phys. Rev. B* **40**, 546 (1989).

[14] L. D. Landau, E. M. Lifshitz, *Statistical Physic, Part 2*, (Pergamon Press, Oxford, 1980) pp. 99-105.

[15] N. W. Ashcroft, N. D. Mermin, *Solid State Physics*, (Saunders College, Philadelphia, 1976) pp. 784-789.

[16] E.g., a 3-level laser has a similar nonequilibrium loop to the one described here.

[17] Kim and Chan's torsional oscillator cell is only 3D for excitations above $n\sim 65$. For temperatures of interest, the peak of the phonon energy distribution is $n\sim 2\times 10^4$, making the 3D energy density expression a good approximation.

[18] M. W. Meisel, *Physica B* **178**, 121 (1992).

[19] J. M Goodkind, *Phys. Rev. Lett.* **89**, 095301 (2002).

**Figure 1: The non-equilibrium loop. A.) In equilibrium at finite temperature below $T_c$, some quasiparticles macroscopically occupy the ground state (lower oval), and the rest (thin line) are spread out among the excited states according to the Bose distribution. B.) Collective scattering of a single phonon by many quasiparticles (upper oval) causes an excited state to be macroscopically occupied. C.) Other scattering processes break up the macroscopically occupied excited state, removing the return path via collective scattering. D.) Quasiparticles return to the ground state via other scattering processes. The resulting steady state distribution is not the Bose distribution.**

**Figure 2: The normal solid to supersolid transition. The non-classical rotational inertia fraction (NCRIF), which is proportional to the number of Bose condensed quasiparticles, is plotted with respect to temperature at three different pressures. A.) P=25 bar. B.) P=41 bar. C.) P=65 bar. The solid circles are the measured values of Kim and Chan. The dashed line is $N_{c0}$, the result for a thermalized Bose gas. The solid line is the non-equilibrium theory assuming constant β, the dotted line assumes β is proportional to the number of unequilibrated quasiparticles. The model constrains the data to lie between the solid and dotted curves, as they do. The**

parameters for the various theoretical curves are: A.) $T_{sa}$=28 mK, $T_{sb}$=120 mK, N=.0072, $T_c$=2 K; B.) $T_{sa}$=14 mK, $T_{sb}$=48 mK, N=.017, $T_c$=1.4 K; C.) $T_{sa}$=19 mK, $T_{sb}$=67 mK, N=.015, $T_c$=0.4 K.

**Figure 3: Proposed $^4$He phase diagram. The normal solid/unequilibrated supersolid line (inverted lambda line) is found by calculating $T_c$ (the solid circles) using Kim and Chan's measured low temperature NCRIF and $m_{eff}$. (We use all the data for which $v_c$ can be determined except for P=25 bar, where the calculation yields $T_c$ ~15 K.) The unequilibrated supersolid/supersolid line is associated with the experimental $T_s$ (the x's), where the data start to conform to Bose statistics. The squares in the unequilibrated supersolid regime correspond to the highest temperatures at which non-classical inertial effects are observed. From the perspective of this framework, these points are not particularly important. The ground state is macroscopically occupied to the right of the points, but to a degree that decreases as $T^{3.5}$.**

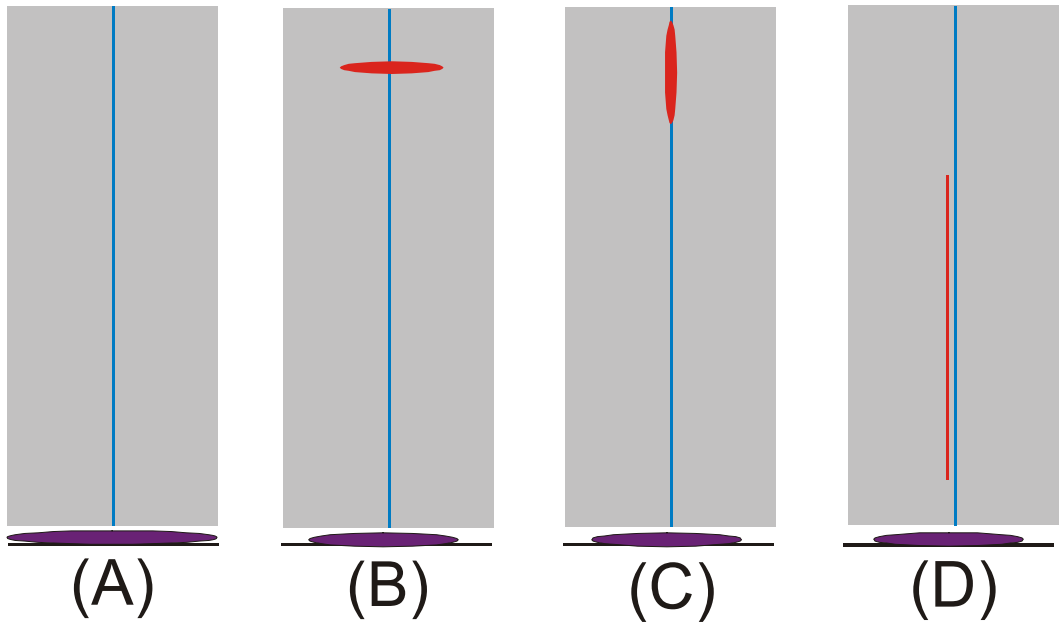

Figure 1

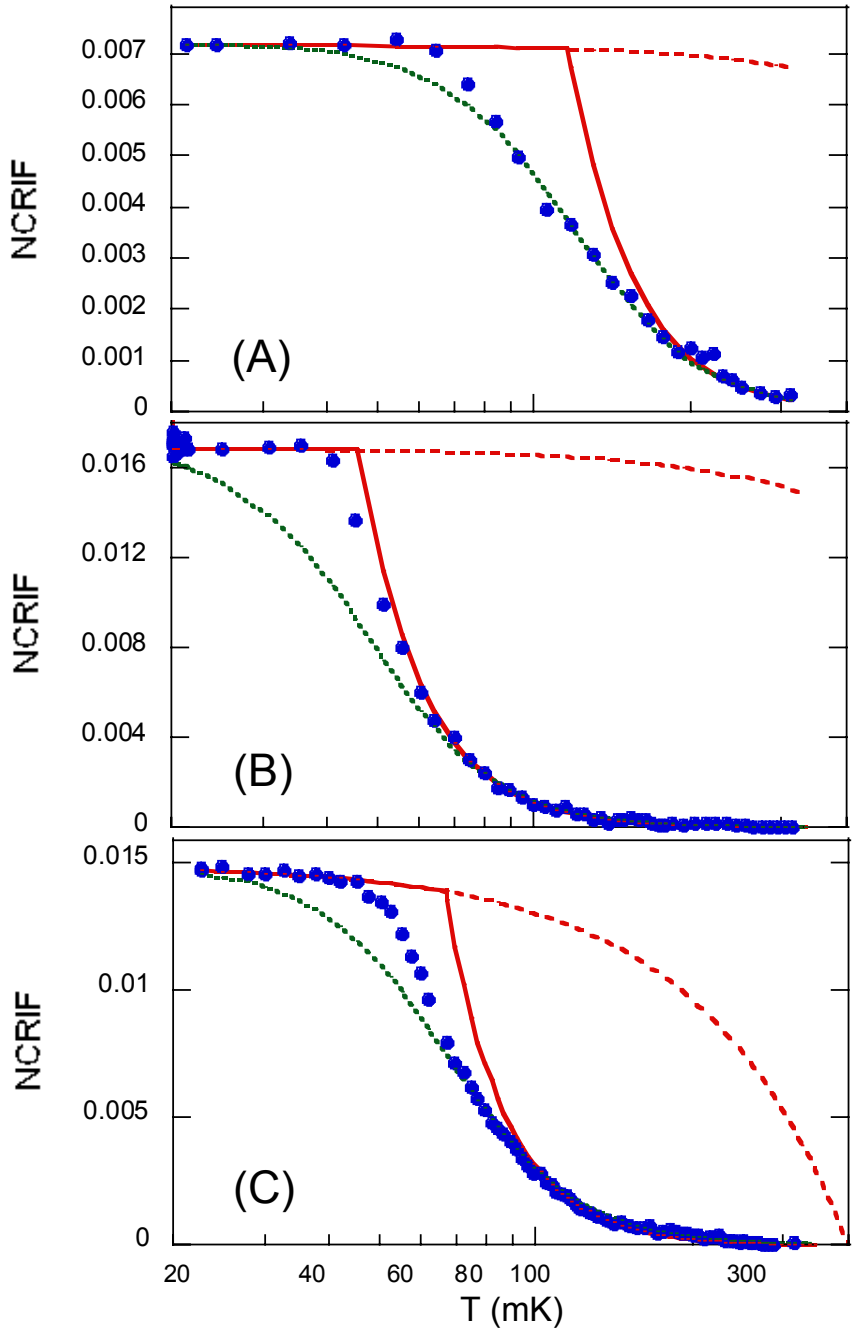

Figure 2

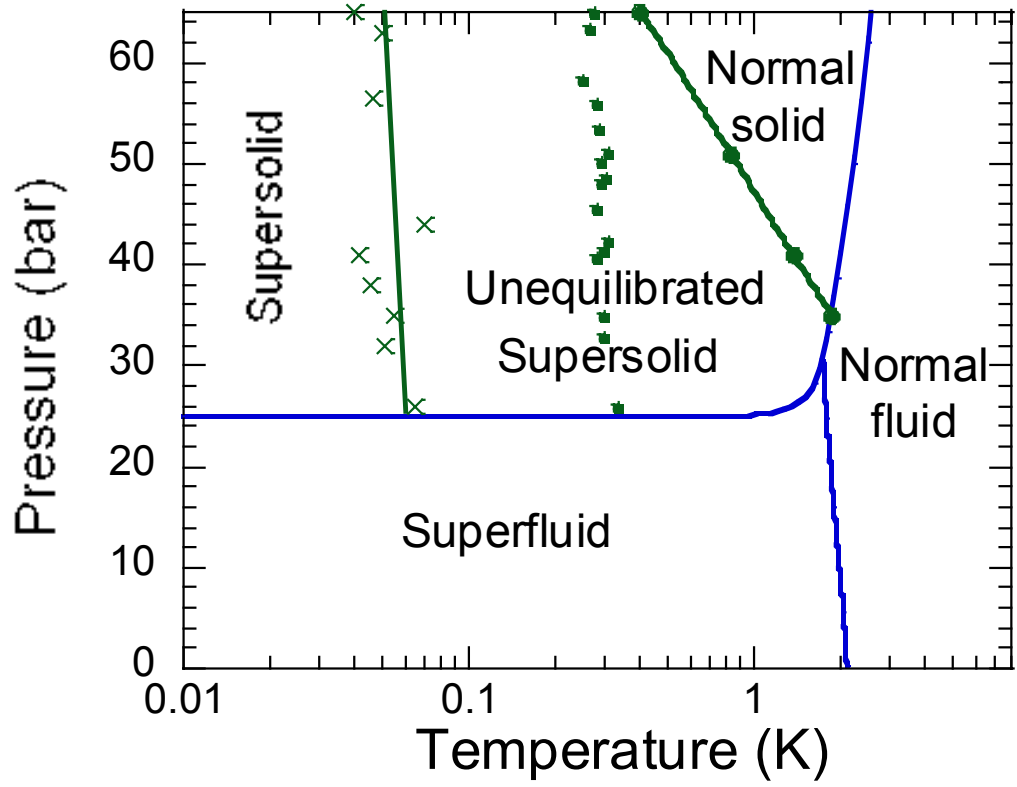

Figure 3